\documentclass[a4paper,french]{rnti}

\usepackage[T1]{fontenc}
\usepackage{color}

\usepackage{url}

\usepackage{graphicx}
\usepackage{amsfonts}
\usepackage{amsmath}
\usepackage{nccmath}
\usepackage{subcaption}
\usepackage{multirow}
\usepackage{multicol}
\usepackage[normalem]{ulem}

\usepackage{amssymb}
\usepackage{amsmath}
\usepackage{graphicx}
\usepackage[colorlinks=true, allcolors=blue]{hyperref}
\usepackage{adjustbox}

\titrecourt{Augmentation de données par interpolation}

\titre{Interpolation pour l'augmentation de données géo-référencées  : Application à la prédiction des adventices de la canne à sucre à La Réunion}

\nomcourt{Fabre Ferber et al.}

\auteur{Frédérick Fabre-Ferber\affil{1}\affil{2},
        Dominique Gay\affil{2}, Jean-Christophe Soulie\affil{2}, 
        Jean Diatta\affil{1},\\
        Odalric-Ambrym Maillard\affil{3},
}

\affiliation{
    \affil{1}LIM EA2525, Université de La Réunion\\
          frederick.fabre@univ-reunion.fr 
          dominique.gay@univ-reunion.fr  
          jean.diatta@univ-reunion.fr \\
    \affil{2}UPR Recyclage et risque, CIRAD\\
          jean-christophe.soulie@cirad.fr 
          frederick.fabre-ferber@cirad.fr\\

    \affil{3}SCOOL, Inria\\
        odalric.maillard@inria.fr
 }
\resume{
L'augmentation de données est une étape cruciale pour le développement de modèles d'apprentissage  supervisés robustes, en particulier lorsqu'il s'agit de jeux de données limités. Cette étude explore des techniques d'interpolation pour l'augmentation de données géo-référencées, avec pour objectif de prédire la présence de Commelina benghalensis L. dans les parcelles de canne à sucre à La Réunion. Compte tenu de la nature spatiale des données et du coût élevé de leur collecte, nous avons évalué deux approches d’interpolation : les processus gaussiens (GPs) avec différents noyaux et le krigeage avec divers variogrammes. 
Les objectifs de ce travail sont triples : (i) identifier quelles méthodes d'interpolation offrent les meilleures performances prédictives pour divers algorithmes de régression, (ii) analyser l'évolution des performances en fonction du nombre d'observations ajoutées, et (iii) évaluer la cohérence spatiale des ensembles de données augmentés. Les résultats montrent que les méthodes basées sur les GPs, en particulier avec noyaux combinés (GP-COMB), améliorent significativement les performances des algorithmes de régression tout en nécessitant moins de données supplémentaires. Bien que le krigeage présente des performances légèrement inférieures, il se distingue par une couverture spatiale plus homogène, un avantage potentiel dans certains contextes.
}
\summary{Data augmentation is a crucial step in the development of robust supervised learning models, especially when dealing with limited datasets. This study explores interpolation techniques for the augmentation of geo-referenced data, with the aim of predicting the presence of Commelina benghalensis L. in sugarcane plots in La Réunion. Given the spatial nature of the data and the high cost of data collection, we evaluated two interpolation approaches: Gaussian processes (GPs) with different kernels and kriging with various variograms. 
The objectives of this work are threefold: (i) to identify which interpolation methods offer the best predictive performance for various regression algorithms, (ii) to analyze the evolution of performance as a function of the number of observations added, and (iii) to assess the spatial consistency of augmented datasets. The results show that GP-based methods, in particular with combined kernels (GP-COMB), significantly improve the performance of regression algorithms while requiring less additional data. Although kriging shows slightly lower performance, it is distinguished by a more homogeneous spatial coverage, a potential advantage in certain contexts.}
\begin{document}
\section{Introduction}
Dans le domaine de l’intelligence artificielle centrée sur les données, l’augmentation de données joue un rôle crucial dans le développement des ensembles d’apprentissage \citep{dcai23}. Ce processus consiste à enrichir le jeu d’entraînement en générant de nouvelles observations sans collecter directement de données supplémentaires. Dans la littérature, deux termes clés émergent : Data Augmentation (DA) et Data Generation (DG). Ces concepts recouvrent plusieurs pratiques, telles que l’augmentation d’observations, la création de nouvelles variables, ou encore le traitement des valeurs manquantes \citep{cui2024tabular}. Dans cette étude, nous nous focalisons sur l’augmentation d’observations, que nous désignerons par data augmentation, dans le contexte de données tabulaires.
Parmi les approches existantes, les méthodes basées sur des règles (rule-based) extraient des motifs spécifiques dans les données pour générer de nouvelles observations \citep{li2021rule}. Les autoencodeurs variationnels (VAE) combinent apprentissage non supervisé et probabilités pour produire de nouvelles données \citep{zhu_s3vae_2020, wan_variational_2017, lotfollahi2020conditional}. D’autres techniques incluent les réseaux antagonistes génératifs (GANs) \citep{ouyang2023missdiff, zheng2022diffusion, kotelnikov2023tabddpm}, les modèles de diffusion \citep{kotelnikov2023tabddpm, ouyang2023missdiff}, et l’apprentissage par renforcement \citep{esnaashari2021automation, li2021rule, yu_data_2022}.
Dans notre cas, nous abordons un problème spécifique lié à l’augmentation de données dans un jeu visant à prédire la présence de l’espèce Commelina benghalensis L. dans des parcelles de canne à sucre sur l’île de La Réunion. La collecte de ces données est chronophage, coûteuse, et nécessite une expertise spécifique pour garantir leur qualité. Ces contraintes limitent considérablement le volume de données disponibles, contrairement à des contextes où les approches évoquées plus haut sont couramment utilisées. De plus, nos données présentent une dimension spatiale, chaque parcelle étant géographiquement localisée sur l’île.
Pour répondre à ces défis, nous explorons des méthodes d’interpolation adaptées aux données géo-référencées, notamment le krigeage et les processus gaussiens (GPs). Le krigeage est une méthode géostatistique qui estime les valeurs des points non échantillonnés en utilisant un variogramme pour modéliser la dépendance spatiale. Cette approche repose sur l’idée que des points proches dans l’espace ont des valeurs corrélées. À l’inverse, les GPs définissent une distribution probabiliste sur les fonctions possibles, permettant une interpolation flexible et généralisée.
L’objectif de cette étude est triple : (i) évaluer si ces approches peuvent améliorer significativement les performances prédictives en augmentant les observations disponibles, (ii) analyser l’évolution des performances en fonction du nombre d’observations ajoutées, et (iii) examiner si ces méthodes préservent la cohérence spatiale des données initiales.
Dans la suite de cet article, nous détaillerons les concepts de krigeage et de GPs en section 2. La section 3 présentera la méthodologie et le protocole expérimental mis en œuvre pour répondre à ces questions. Les résultats seront analysés et discutés en section 4, avant de conclure en section 5.
\section{Interpolation par Krigeage et Processus Gaussiens} 

L'interpolation se réfère à la capacité d'un modèle à prédire des valeurs pour des points de données situés entre les observations existantes dans l'ensemble d'entraînement. Contrairement à l'extrapolation, qui concerne les prédictions en dehors de la plage des données connues, l'interpolation se concentre sur les prédictions à l'intérieur de cette plage. Elle joue un rôle crucial dans de nombreux algorithmes, en permettant d'estimer les valeurs continues d'une variable cible en fonction des variables d'entrée. Pour, un jeu de données d’entraînement $D = \{ (x_i,y_i) | i= 1,2,\dots, n \}$ et $y_i = f(x_i)$ un scalaire pour une fonction inconnue $f$ l'on souhaite interpoler la valeur $y^\star $ d'un point de test $x^\star$ à travers $f$.  Le krigeage et les GPs sont deux variantes de la même méthode d'interpolation avec le krigeage qui est le terme traditionnellement utilisé en géo-statistiques et GP pour le machine-learning. Bien qu'ils désignent la même famille d'algorithmes ils présentent cependant quelques différences notamment sur la fonction utilisée pour exprimer la co-variance entre les données (l'un les noyaux et l'autre les variogrammes) et également sur la méthode d'estimation utilisée pour prédire $y^\star$.

\paragraph{Krigeage} 

Le krigeage \citep{cressie1990origins} nommé d'après le géostatisticien sud-africain Danie Krige, est une méthode d'interpolation géostatistique largement utilisée pour estimer les valeurs de points non échantillonnés à partir de données spatiales existantes. Il repose sur la théorie des variables régionalisées et utilise des variogrammes pour modéliser la structure de dépendance spatiale entre les points de données. 
Pour une variable d'intérêt à une position spatiale $x^\star$, le but du krigeage est de prédire $f(x^\star)$ en utilisant une combinaison linéaires des valeurs observées $y_i$ pour les points $x_1 , \dots x_N$ :

$$ f(x^\star) - \mu(x^\star) = \sum_{i=1}^N \lambda_i [ y_i - \mu(x_i)] $$

où \( \lambda_i \) sont les poids attribués à chaque observation $x_i$, $\mu(x_i)$ la moyenne des points d'observations de manière à minimiser l'erreur quadratique moyenne de prédiction. Plusieurs types de krigeage existent en fonction de quelles hypothèses on peut émettre sur la moyenne $\mu(x^\star)$, le krigeage simple suppose que cette moyenne est connue, le krigeage ordinaire suppose que la moyenne est inconnue et elle est donc estimée localement en fonction des points d'observations et le co-krigeage utilisent plusieurs variables corrélés en plus des variables spatiales pour améliorer les prédictions.
Pour calculer les poids  $\lambda_i $, on utilise un variogramme (ou semi-variogramme ) qui décrit la variation de $f(x)$ en fonction de la distance entre deux points \citep{oliver2015basic}.
Le variogramme est crucial dans le krigeage car, il informe sur les pondération des observations pour estimer les valeurs en des points non échantillonnés :  

$$\begin{bmatrix}
\lambda & \dotsc  & \Sigma_{N1}\\
\vdots  & \Sigma_{ij} & \vdots \\
\Sigma_{1N} & \dotsc  & \Sigma_{NN}
\end{bmatrix} \ \begin{bmatrix}
\lambda _{1}\\
\vdots \\
\lambda _{N}
\end{bmatrix} \ =\ \begin{bmatrix}
\Sigma_{x^\star 1}\\
\vdots \\
\Sigma_{x^\star N}
\end{bmatrix}$$

où $\Sigma_{i,j}$ est la covariance entre les points observés $x_i,x_j \in X$ avec $i,j \in \{1,\dots,N \}$ et $\Sigma_{x^\star i}$ la covariance entre le points $x^\star$ à interpoler et un point d'observation $x_i$.

\paragraph{Régression par Processus Gaussien}
 Les processus Gaussiens (GP) \citep{williams2006gaussian} sont des modèles non paramétriques défini comme une collection de variables aléatoires $X$ telles que pour tout ensemble fini de points \( x_1, x_2, \ldots, x_n \), le vecteur \( (f(x_1), f(x_2), \ldots, f(x_n)) \) suit une distribution normale multivariée. Un GP est entièrement spécifié par sa fonction de moyenne \( m(X) = \mathbb{E}[f(X)] \) et sa fonction de covariance \( k(x, x') = \text{Cov}(f(x), f(x')) \). Il estime la valeur $f(x^\star)$ tel que :  

$$ f(x^\star) \sim \mathcal{GP}(m(x) , k(x,x'))$$

Au fur et à mesure que des observations sont faites, cet a priori peut être mis à jour de manière séquentielle pour devenir une distribution a posteriori de la véritable fonction représentant la variable cible d'intérêt. Pour la régression le GP va agir comme une interpolation en calculant la distribution jointe entre les données d’entraînement et le ou les points à prédire  : 

 $$ \begin{bmatrix}
\mathbf{f}\\
\mathbf{f_\star}
\end{bmatrix} \ \sim \ \mathcal{N} \ \left(\begin{bmatrix}
\mu\\
\mu_\star
\end{bmatrix} \ ,\ \begin{bmatrix}
K & K\star\\
K\star^T & K \star \star 
\end{bmatrix} \ \right)$$  
avec $\mathbf{f}$ contenant l'ensemble des valeurs cibles $\{y_1,\dots, y_n\}$ de l'ensemble d’entraînement, $\mathbf{f \star}$ les valeurs prédites sur l'ensemble de test, $\mu$ la moyenne sur l'ensemble d’entraînement, $\mu \star$ la moyenne présumée sur l'ensemble de test, $K$ la covariance entre les points d’entraînement, $K\star$ la covariance entre les points d’entraînement et les points de test et $K \star \star $ la covariance entre les points de test.

\section{Méthode proposée}

L'interpolation dans le cas de données géo-référencées peut être vue comme une augmentation de données. En effet, les algorithmes d'interpolation bien adaptés à un cadre spatial semblent une solution particulièrement adaptée dans ce cas de figure. Ces méthodes d'interpolation peuvent être utilisée pour alors augmenter le nombre de données lorsque celles ci sont en nombre limité. 
Nous choisissons d"utiliser l'interpolation par  krigeage et régression par processus gaussiens présentés dans la section précédente pour augmenter le nombre de données disponible dans une base de données relatives à l'agriculture ayant pour but de prédire le recouvrement de l'espèce adventice \textit{Commelina benghalensis L.} (COMBE) sur des parcelles de canne à sucre à La Réunion. Trois questions peuvent alors se poser : (i) Quelle méthode d'interpolation pour augmenter le nombre de données semble la plus efficace en terme de performance prédictive, (ii) à partir de quel seuil de points ajoutés la performance prédictive semble converger et (iii) comment évolue la distribution de l'espèce (en terme de densité de recouvrement) pour une méthode d'interpolation comparée à la distribution sur le jeu de données sans augmentation ? 

\paragraph{Jeu de données} 

Nous utilisons un ensemble de données comprenant des données d'enquête floristique réelles géo-référencées à La Réunion, qui concernent la présence de l'adventice \textit{Commelina benghalensis L.}  sur des parcelles de canne à sucre \citep{fabre-ferber_dataset_2021,laine2024impact} . Ce jeu de données reflète fidèlement les défis courants dans des contextes de données réalistes et complexes, caractérisés par des observations limitées. Le jeu de données est constitué de 745 observations et comporte 8 variables, à la fois continues et catégorielles résumé dans la table \ref{tab:data}. Les variables comme le lieu de la parcelle ont été retirées ne servant pas à l'interpolation. 
Pour la variable d'intérêt (le recouvrement de l'espèce) est une variable continue allant de 0 à 100, 0 représentant un recouvrement nul sur une parcelle et 100 représentant un recouvrement total de l'espèce \textit{Commelina benghalensis L.}.

\begin{table}[htbp!]
    \centering
    \begin{tabular}{c c c}
     \hline 
      Nom  & Type  & Plage de valeurs\\
       \hline 
       Longitude  & Continue & $[55.23-55.83]$\\
       Latitude  & Continue & $[-21.39-21.88]$ \\
       Altitude  & Continue & $[0-950]$\\
       Température Moyenne  & Continue & $[18-28]$\\
       Précipitation  & Continue & $[0-1400]$ \\
       Mois & Catégrorielle & $[1-12]$ \\
       Année  & Catégorielle & $[2002-2024]$\\
       Luminance  & Continue & $[900-1890]$ \\
       \hline
    \end{tabular}
    \caption{Descriptif des différentes variables présentes dans le jeu de données pour le recouvrement de l'espèce \textit{Commelina benghalensis L.}}
    \label{tab:data}
\end{table}

\paragraph{Méthodes d'interpolation}

Dans un cas classique d'interpolation, où l'on souhaite estimer la valeur d'un point géo-référencé $x^\star$ à partir des points observés, celui est défini par ses coordonnées dans un système de projection donné (dans notre cas, latitude et longitude). Ces coordonnées peuvent être directement obtenues grâce à la connaissance du système de projection. Toutefois, le problème se complexifie lorsque l'on souhaite intégrer d'autres variables auxiliaires, comme l'altitude ou la pluviométrie, qui ne peuvent pas être directement déduites du système de projection. Une approche simple consisterait à assigner une valeur moyenne ou une distribution calculée à partir des points observés. Cependant, ces hypothèses peuvent introduire des écarts significatifs par rapport à la réalité, compromettant ainsi la précision de l'interpolation.

Dans notre approche, lorsque nous créons un point $x^\star$ à interpoler pour augmenter le nombre de point dans la base d'apprentissage, nous récupérons les différentes variables auxiliaires à l'aide du service Meteor\footnote{https://smartis.re/METEOR}, qui nécessite une date et une localisation. Pour garantir une répartition temporelle équilibrée, nous choisissons des données couvrant toutes les périodes de l'année sans surreprésentation de certains mois. De plus, nous restreignons les points à interpoler à des zones où la probabilité de trouver des parcelles de canne à sucre est élevée, évitant ainsi des régions non pertinentes. Concernant les méthodes d'interpolation, nous utilisons plusieurs noyaux pour les processus gaussiens : un noyau linéaire (équation \ref{eq:lin}), un noyau RBF (équation \ref{eq:rbf}) et un noyau quadratique (équation \ref{eq:poly}). Nous appliquons également une méthode de recherche de combinaisons de noyaux décrite dans \cite{duvenaud2014kernel}, où les noyaux sont construits à l'aide de divers opérateurs :

$$ \mathcal{S} \rightarrow \mathcal{S+B} $$ 
$$\mathcal{S} \rightarrow \mathcal{S\times B} $$
$$\mathcal{B} \rightarrow \mathcal{B'} $$ avec $\mathcal{S}$ qui représente n'importe quelle sous-expression de noyau et $\mathcal{B}$ un noyau de base. Les noyaux de base sont ceux décrits précédemment et le critère utilisé pour choisir un noyau optimal est le critère d'information bayesien (BIC) \cite{schwarz1978estimating}. Le co-krigeage, une variante du krigeage classique, est utilisé pour intégrer d'autres variables explicatives dans l'analyse. Plusieurs modèles de variogrammes sont employés, notamment le variogramme linéaire (équation \ref{eq:lin_kr}), exponentiel (équation \ref{eq:exp}), gaussien (équation \ref{eq:gau}) et sphérique (équation \ref{eq:sphe}).

\begin{align}
    k(x, x') &= x \cdot x' \label{eq:lin} \\
    k(x, x') & = \exp\left(-\frac{\|x - x'\|^2}{2\sigma^2}\right) 
 \label{eq:rbf} \\
    k(x, x') &= (x \cdot x' + c)^2  \label{eq:poly}
\end{align}

\begin{align}
    \gamma(h) &= C_0 + bh \label{eq:lin_kr} \\
   \gamma(h) &= C_0 + C \left(1 - \exp\left(-\frac{h}{a}\right)\right) 
 \label{eq:exp} \\
    \gamma(h) &= 
\begin{cases}
    C_0 + C \left( \frac{3h}{2a} - \frac{h^3}{2a^3} \right), & \text{si } 0 \leq h \leq a \\
    C_0 + C, & \text{si } h > a
\end{cases} \label{eq:sphe} \\ 
\gamma(h) &= C_0 + C \left(1 - \exp\left(-\frac{h^2}{a^2}\right)\right) \label{eq:gau}
\end{align} 

avec $C_0$ l'effet pépite, $C$ le palier, $h$ la distance et $a$ un paramètre d'échelle. 

\paragraph{Protocole}

Le protocole mis en oeuvre dans cette étude vise à répondre aux questions posées en début de section. Nous évaluons la performance prédictive de plusieurs algorithmes de régression, mesurée en termes d'erreur quadratique moyenne (MSE). Les algorithmes testés incluent la régression linéaire (LR), la régression Ridge (RR), le Support Vector Machine Regressor (SVR), les forêts aléatoires (RF), le Gradient Boosting Regressor (GB), $k$-plus proches voisins (KNN) et les réseaux de neurones (MLP). Ces performances sont comparées entre le jeu de données initial et le jeu augmenté par différentes méthodes d'interpolation (tab \ref{tab:my_label}).  Trois expérimentations sont menées : (i) L'évaluation des performances sur un ensemble de test représentant 30\% des données, en comparant les différents algorithmes de régression appliqués sur le jeu de données d'origine et celui augmenté avec 200 points supplémentaires. (ii) L'analyse des performances de l'algorithme le plus performant issu de la première expérimentation, en fonction d'un nombre croissant de points ajoutés $\{0,50,\dots,300\}$, où 0 correspond au jeu de données d'origine. (iii) Comparaison de cartes de densité en termes de taux de recouvrement, entre celles obtenues avec le jeu de données initial et celles générées par les différentes méthodes d'interpolation où nous rajoutons 300 nouveaux points. 
Le choix des paramètres des kernels pour les GPs est effectué par descente de gradient du logarithme de la vraisemblance avec la librairie \texttt{GPy}\footnote{https://gpy.readthedocs.io/en/deploy/}. Pour les paramètres des variogrammes, ne disposant pas de méthodes automatiques ceux ci sont choisis selon un ensemble de paramètres en minimisant la MSE avec la librairie \texttt{PyKrige} \footnote{https://geostat-framework.readthedocs.io/projects/pykrige/en/stable/}. Les hyper-paramètres des algorithmes de régression sont laissés par défaut et sont issus de la librairie \texttt{scikit-learn} \cite{scikit-learn}.

\begin{table}[htbp!]
    \centering
    \begin{tabular}{c c}
     \hline 
      Méthode  & Acronyme \\
       \hline \\
       Processus Gaussien avec kernel linéaire  & GP-LIN \\
       Proccessus Gaussien avec kernel polynomial & GP-POLY \\
       Processus Gaussien avec kernel RBF & GP-RBF \\
       Processus Gaussien avec combinaison de kernels & GP-COMB\\
       Krigeage avec variogramme linéaire & COKR-LIN \\
       Krigeage avec variogramme exponentiel & COKR-EXP \\
       Krigeage avec variogramme gaussien & COKR-GAU \\
       Krigeage avec variogramme sphérique & COKR-SPHE \\
       \hline
    \end{tabular}
    \caption{Méthodes d'interpolation utilisées}
    \label{tab:my_label}
\end{table}

\section{Resultats et interprétation} 

\begin{figure}[th!]
    \centering
    \includegraphics[width=0.9\linewidth]{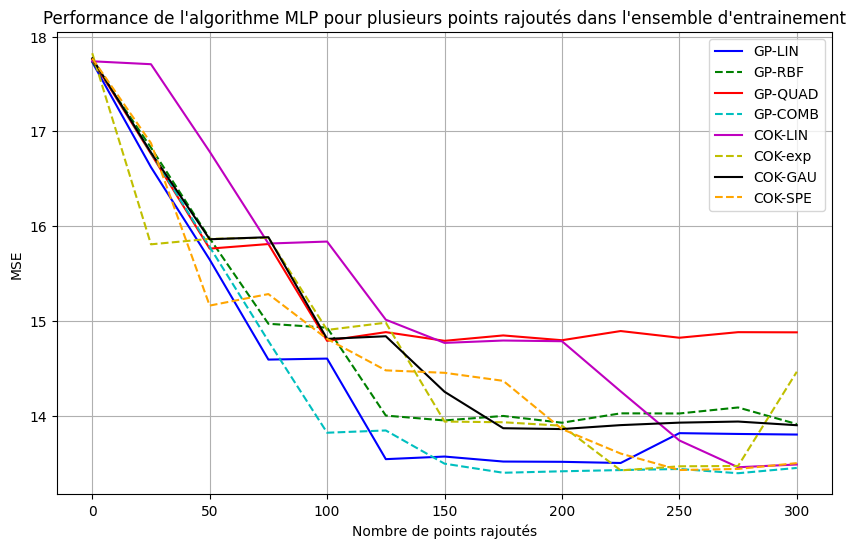}
    \caption{Performance en terme de MSE de l'algoritme MLP pour 0 à 300 points rajoutés par différentes techniques d'interpolation. }
    \label{fig:enter-label}
\end{figure}

\begin{figure}[htbp!]
    \centering
    \includegraphics[width=0.99\linewidth]{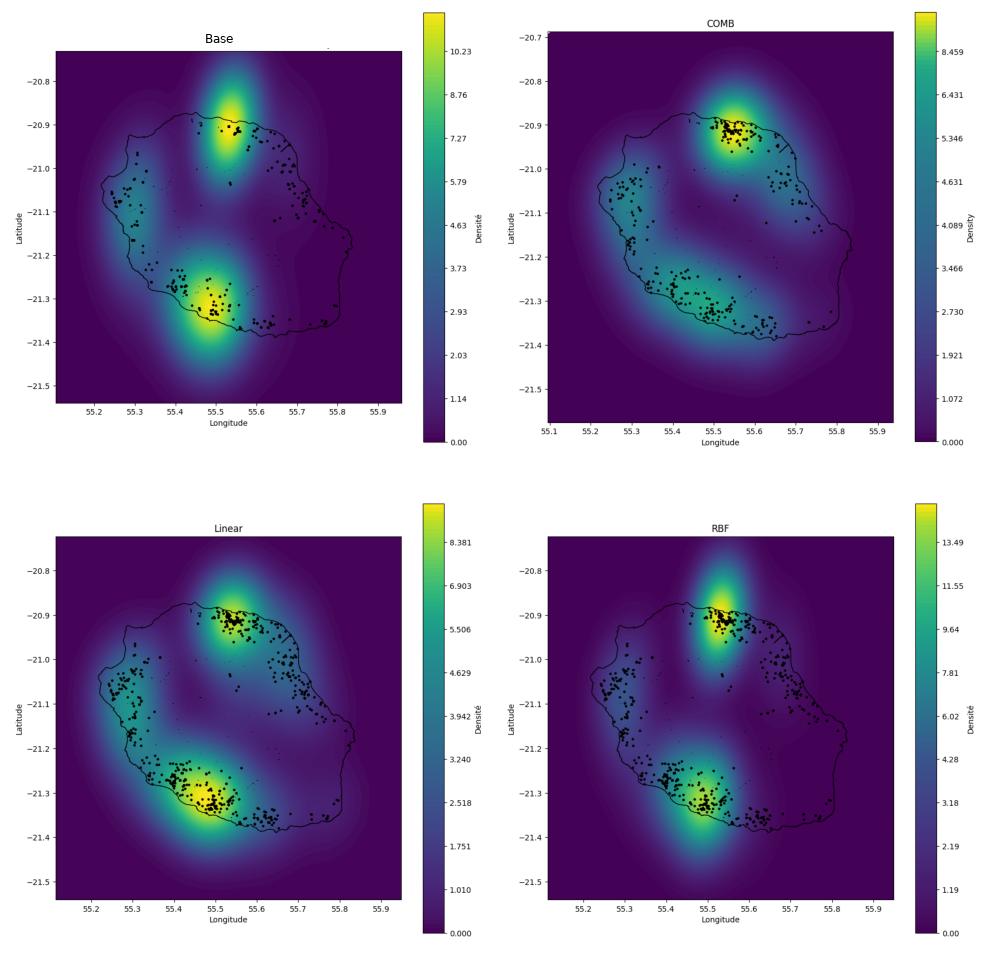}
    \caption{Carte de densité de recouvrement pour l'espèce \textit{Commelina benghalensis L.} (COMBE) pour le jeu de données de base (Base) et les jeux de données augmentés par 3 méthodes d'interpolation GP-COMB,GP-LIN et GP-RBF. Nous choisissons de ne pas afficher COK-EXP et COK-SPHE, la différence de recouvrement étant minime.}
    \label{fig:density}
\end{figure}

    Cette section présente et analyse les résultats des différentes augmentations de données via les techniques d'interpolations décrites précédemment. 
    Les résultats de la table 2 présentent les erreurs quadratiques moyennes (MSE) pour divers algorithmes de régression appliqués à des jeux de données augmentés via différentes techniques d'interpolation.  La figure \ref{fig:enter-label} montre l'évolution de la performance pour un algorithme particulier plusieurs nombre de points rajoutés par les différentes interpolations. Enfin, la figure \ref{fig:density} montre des carte de densités faite avec le jeu de données de base et les jeux de données augmentés par les différentes techniques d'interpolation. 
\begin{table}[htbp!]

    \label{tab:results}
    \begin{adjustbox}{width=\textwidth,center}
    \begin{tabular}{lccccccccc}
        \hline
        
        \textbf{Modèle} & \textbf{Base} & \textbf{GP-RBF} & \textbf{GP-LIN} & \textbf{GP-QUAD} & \textbf{GP-COMB} & \textbf{CoK-LIN} & \textbf{CoK-EXP} & \textbf{CoK-GAU} & \textbf{CoK-SPHE} \\
        \hline
        \textbf{LR} & 18.80 & 13.68 & 13.71 & \textbf{13.67} & 14.57 & 14.58 & 14.78 & 14.66 & 14.56 \\
        \textbf{RR} & 14.84 & 13.68 & 13.71 & \textbf{13.67} & 13.57 & 14.57 & 14.58 & 14.66 & 14.56 \\
        \textbf{SVR} & 16.80 & 14.09 & \textbf{13.74} & 13.93 & 14.98 & 14.87 & 15.12 & 15.11 & 15.11 \\
        \textbf{RF} & 23.83 & 14.05 & 14.05 & 14.01 & \textbf{13.35} & 13.35 & 13.62 & 13.53 & 13.55 \\
        \textbf{BG} & 36.45 & 13.65 & 13.55 & 13.57 & \textbf{13.27} & 13.34 & 13.28 & 13.28 & 13.34 \\
        \textbf{KNN} & 16.30 & 14.44 & 14.55 & 14.44 & \textbf{13.17} & 13.18 & 13.21 & 13.17 & 13.17 \\
        \textbf{MLP} & 17.17 & 13.47 & 13.55 & 13.54 & \textbf{13.41} & 13.49 & 13.47 & 13.44 & \textbf{13.38} \\
        \hline
        
    \end{tabular}
    \end{adjustbox}
        
        \caption{Résultats des modèles de régression avec différentes méthodes d'interpolation pour 200 points générés}
\end{table}
    
    \paragraph{Évaluation} Les résultats montrent que l’augmentation des données via Processus Gaussiens et Krigeage améliore systématiquement les performances des modèles de régression comparées à Base. Parmi les techniques testées, GP-COMB s’impose comme la méthode la plus performante pour RF, GB et K-NN, montrant qu'elle semble efficace pour générer des nouvelles données servant bien à la performance. GP-QUAD se distingue pour LR et RR, en modélisant efficacement les relations à la fois linéaires et non linéaires, ce qui le rend particulièrement adapté à ces modèles.
    Pour l'interpolation par krigeage, les variogrammes Co-K-SPHE et Co-K-EXP produisent de bons   résultats, notamment pour MLP et K-NN. Toutefois, ils montrent une compatibilité limitée avec certains modèles comme SVR, où les performances restent inférieures à celles des processus gaussiens. Les résultats révèlent également que l’utilisation du jeu de données sans augmentation entraîne une baisse notable des performances, ce qui souligne la valeur des approches d’augmentation, en particulier dans les contextes où les données initiales sont limitées.
    En résumé, GP-COMB se distingue comme la méthode la plus efficace et polyvalente, offrant des améliorations significatives pour de nombreux modèles mais particulièrement pour MLP qui en moyenne sur toutes les méthodes d'interpolation à les meilleures performances .Toutefois, le krigeage avec l'ensemble des variogrammes demeure une alternative intéressante, notamment pour les modèles comme MLP, et peut être privilégiée selon le contexte et les caractéristiques des données. Nous choisissons pour la suite de ces expérimentations de garder l'algorithme MLP.

\paragraph{Performance en fonction du nombre de points rajoutés}

L'analyse de l'évolution de la MSE en fonction du nombre de points ajoutés a mis en évidence des comportements distincts selon les méthodes d'interpolation. GP-COMB et GP-LIN se sont démarqués par une convergence rapide vers une erreur minimale, atteignant un plateau vers 200 points ajoutés. Ceci montrent que ces méthodes convergent plus vite et on a donc moins la nécéssité de rajouter plus de points. En revanche, les autres méthodes, bien qu'efficaces, ont généralement nécessité un plus grand nombre de points pour atteindre un plateau. COK-LIN, par exemple, a montré des performances comparables aux meilleurs GP mais avec une convergence plus lente. Le processus gaussien avec noyau quadratique (GP-QUAD) s'est distingué négativement, son erreur stagnant à un niveau significativement supérieur aux autres méthodes. 
Globalement on observe que toutes les méthodes augmentent la performance de prédiction et qu'elles stagnent quasiement toute vers une même erreur quadratique moyenne ormis pour GP-RBF,GP-QUAD et COK-GAU malgré qu'ils soient tous très proche. En conclusion, l'ajout de points supplémentaires améliore généralement la précision du modèle MLP. Les GPs avec noyau linéaire et la combinaison de noyaux, ainsi que le co-krigeage avec variogramme linéaire et sphérique, se sont avérés particulièrement efficaces pour réduire l'erreur en fonction du nombre de points mais GP-COMB et GP-LIN semblent favorisés stagnant plus vite vers 150-200 points ajoutés.

\paragraph{Carte de densité}

\begin{figure}
    \centering
    \includegraphics[width=0.5\linewidth]{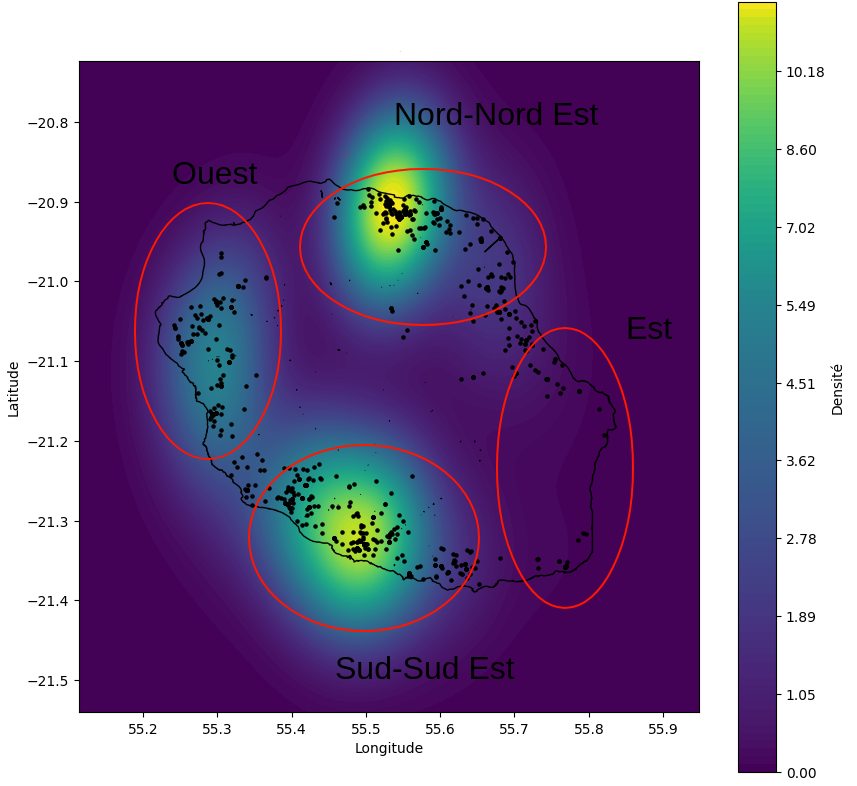}
    \caption{Zones importantes de La Réunion pour l'apprition de l'espèce  \textit{Commelina benghalensis L.}}
    \label{fig:exemple}
\end{figure}

Dans cette analyse, nous comparons des cartes de densité de recouvrement d'espèces générées à partir du jeu de données d'origine et des différentes méthodes d'interpolation, notamment GP-LIN, GP-RBF, GP-COMB, ainsi que les variantes de krigeage COK-SPHE et COK-EXP (table \ref{fig:density}). L'objectif est d'évaluer dans quelle mesure ces méthodes d'augmentation influencent la répartition spatiale de l'espèce dans différentes zones de l'île : Nord-Nord Est, Ouest, Sud-Sud Est (figure \ref{fig:exemple}). 
En plus des différentes cartes, nous calculons la différence moyenne de recouvrement entre les points d'une zone du jeu de données d'origine et les zones des jeux de données avec les points rajoutés (table \ref{tab:diff}).

\begin{table}[]
    \centering
    \begin{tabular}{cccccc}
        \hline
       & GP-COMB & GP-LIN & GP-RBF & COK-SPHE & COK-EXP \\
       \hline
         Nord-Nord Est & \textbf{+1.1} &  \textbf{+1.27} & \textbf{+1.9}  & -0.2 & -0.2\\
         Est &  \textbf{+1.56} & \textbf{+1.32} & -0.4 & +0.1 & +0.1  \\
         Ouest & +0.2 & +0.2 & \textbf{-0.8} & -0.14 & -0.14 \\
         Sud-Sud Est & \textbf{-1.2} & -0.24 & \textbf{+0.8}  &  -0.1 & -0.1  \\
         \hline
    \end{tabular}
    \caption{Difference de recouvrement moyens entre les points du jeu de données sans augmentation et les jeux de données aumgentés par les différentes techniques d'interpolation. Les différences signicatives sont en gras. }
    \label{tab:diff}
\end{table}

Les méthodes GP-COMB et GP-LIN présentent des similitudes en accentuant les contrastes dans les zones Nord et Est, où les zones de recouvrement sont plus étendues par rapport à la carte de base ce qui explique la valeur plus haute en moyenne de recouvrement. Cependant, GP-COMB à tendance à moins représenter l'espèce dans le sud. GP-RBF, quant à lui, produit des résultats plus contrastés. Il exagère fortement les valeurs dans les zones Nord et Sud (négativement), créant des pics de recouvrement très prononcés dans le nord et un peu moins dans le Sud. Bien qu'il reste cohérent dans la zone de l'Ouest comme toutes les autres méthodes il sous re-présente la zone Est de l'ile avec une différence negative. Les méthodes COK-SPHE et COK-EXP offrent une vision plus homogène de la distribution spatiale, avec des variations beaucoup moins marqués que les autres méthodes.

\paragraph{Résumé des résultats} 
Pour résumer ces résultats, en termes de performance prédictive les modèles GP-COMB et GP-LIN se distinguent mieux des modèles de krigeage sur l'ensemble de validation. De plus, ces méthodes à noyaux atteignent une convergence plus rapide avec un nombre de points réduit. Cependant, l'analyse de la distribution spatiale du recouvrement révèle des différences notables. Les méthodes de krigeage maintiennent une distribution plus homogène du recouvrement sur l'ensemble de l'île, tandis que les méthodes à noyaux tendent à généraliser davantage le recouvrement entre les différentes zones. Cette différence pourrait être liée à la nature plus flexible des noyaux utilisés dans les méthodes GP. Néanmoins, une évaluation approfondie dans un contexte agronomique est nécessaire pour déterminer si cette généralisation est bénéfique ou non.

\section{Conclusion} 

Dans cette étude, nous avons exploré diverses techniques d'interpolation pour l'augmentation de données géo-référencées, en évaluant leur efficacité à travers plusieurs algorithmes de régression. Nous avons également analysé leur impact sur la répartition de l'espèce \textit{Commelina benghalensis L.}. Deux principales approches ont été examinées : les processus gaussiens (GP) avec différents noyaux et le krigeage utilisant divers variogrammes. Cette recherche avait pour objectifs (i) d’identifier les méthodes offrant les meilleures performances sur la validation, (ii) d’analyser l’évolution des performances d’un algorithme en fonction du nombre de points ajoutés, et (iii) de comprendre l’influence de ces approches sur la répartition spatiale estimée par rapport au jeu de données initial.
Les résultats indiquent que certaines méthodes, notamment GP-COMB et GP-LIN, se distinguent en améliorant significativement les performances des algorithmes de régression tout en nécessitant moins de points pour atteindre une convergence optimale. Les méthodes de krigeage, bien que globalement moins performantes et présentant une convergence plus lente, offrent un recouvrement spatial de l’espèce plus homogène. En revanche, les approches basées sur les noyaux tendent à produire des modèles plus généralistes sur l’ensemble de l’île, un résultat qui reste à approfondir pour en évaluer pleinement les implications.
Ces conclusions ouvrent plusieurs perspectives. D’une part, il serait pertinent d’appliquer ces techniques à d’autres jeux de données géo-référencées afin de vérifier leur robustesse et leur applicabilité. D’autre part, une adaptation de ces méthodes d’interpolation à des scénarios multi-labels constitue une piste intéressante. En effet, l’étude actuelle s’est concentrée sur l’augmentation des données pour une seule espèce (une variable cible), mais l’extension à plusieurs variables, simultanément, pourrait élargir considérablement le champ d’application de ces approches.

\bibliographystyle{rnti}
\bibliography{main}

\Fr
\end{document}